# Two tales of science technology linkage: Patent in-text versus front-page references

Jian Wang* and Suzan Verberne

Leiden University

*Corresponding author:

j.wang@sbb.leidenuniv.nl

15 March 2021

**ABSTRACT**

There is recurrent debate about how useful science is for technological development, but we know little about what kinds of science are more useful for technology. This paper fills this gap in the literature by exploring how the value of a patent (as measured by patent forward citations and the stock market response to the issuing of the patent) depends on the characteristics of the scientific papers that it builds on, specifically, basicness, interdisciplinarity, novelty, and scientific citations. Using a dataset of 33,337 USPTO biotech utility patents and their 860,879 in-text references to Web of Science journal articles, we find (1) a positive effect of the number of referenced scientific papers, (2) an inverted U-shaped effect of basicness, (3) an insignificant effect of interdisciplinarity, (4) a discontinuous and nonlinear effect of novelty, and (5) a positive effect of scientific citations for patent market value but an insignificant effect on patent citations. In addition, in-text referenced papers have a higher chance of being listed on the front-page of the same patent when they are moderately basic, less interdisciplinary, less novel, and more highly cited. Accordingly, using front-page reference yields substantially different results than using in-text references.

**Keywords:** Science technology linkage, basicness, interdisciplinarity, novelty, citations, text analytics

# 1. Introduction

How science and technology feed into each other is a long-standing research question, and scholars have contributed much insight into the complex interaction between science and technology (Brooks, 1994; Colyvas et al., 2002; Nelson, 1959; Tijssen, 2018). However, it is still debated whether science is useful for technological development. On the one hand, many argue that although science is remote from the marketplace and has little direct commercial value, it may lead to broad, revolutionary, and unexpected applications (Bush, 1945; Nelson, 1959; Partha & David, 1994; Stephan, 1996). Furthermore, the fundamental understanding resulting from scientific research provides a map for technological innovation and improves the innovation process (Cassiman, Veugelers, & Zuniga, 2008; Fleming & Sorenson, 2004). On the other hand, others argue that science and technology follow distinct logics, which leads to conflicting beliefs and approaches for problem-solving, and as a result, scientific research has little value and can even be detrimental for technological development (Ali & Gittelman, 2016; Gittelman & Kogut, 2003; Nelson, 2003).

These two conflicting views about the science-technology-linkage are also reflected in recurrent debates in science policy regarding whether the government should support science and how public science should be organized. Following Bush (1945)'s vision, the post-WWII science policy has offered scientists abundant government support and autonomy for pursuing basic scientific research without considerations of practical ends, under the expectation that the resulting science will eventually generate practical value. However, policymakers and scientists are increasingly under the pressure to justify the public investment in science, and science is called on to make more direct and active contributions to the economy and society (Etzkowitz & Leydesdorff, 2000; Gibbons, 1994; Lubchenco, 1998). Universities have been mandated to perform the third mission of serving the economy and society (in addition to their traditional missions of teaching and research), and various science policies have been devised to stimulate academic scientists' industry engagement, technology transfer, and entrepreneurial activities (Bozeman, 2000; Bozeman, Rimes, & Youtie, 2015; Perkmann et al., 2013). However, academic scientists are still divided regarding their positions towards this third mission: Some have fully embraced the commercial logic, while others are still strongly committed to the traditional academic logic (Johnson, 2017). In today's science policy discussions, some still echo Bush's vision (National Academies, 2020), while others call for a more fundamental restructuring of the science system (Sarewitz, 2016; Shneiderman, 2018).

A gap in the theoretical and policy debate is that we lack a systematic understanding about what types of scientific research are particularly useful for technological development. Science is heterogenous, and certain types of scientific outputs may contribute to technological development disproportionally more than others. Exploring the heterogeneity in science may offer clues for reconciling the two competing theories about the science-technology-linkage. Therefore, this paper asks the question: What types of scientific output are more useful for technological development, more specifically, how the value of a patent depends on the characteristics of scientific papers underlying the patented technology, in terms of basicness, interdisciplinarity, novelty, and scientific citations? Answering this question will also inform science policy in two ways: First, whether academic scientists' self-governed pursuit of "good" science (in scientific standards) might be at odds with the policy goal of making science more "useful" (for innovation, economy, and society) such that policy intervention is necessary. Second, it is important to know what kinds of science should be supported to achieve the policy goal of "useful" science.

One methodological challenge confronting studies of science-technology-linkage is the lack of large-scale empirical data. There is a rising interest in analyzing patent references to the scientific literature since the seminal work of Narin and Noma (1985). References in patents to scientific papers provide a paper trail of the knowledge flow from science to technology and therefore have enabled a fruitful and vibrant research area. However, the current practice relies almost exclusively on patent front-page references but neglects the more difficult patent in-text references. Some recent studies have suggested that patent in-text and front-page references are generated from different processes and have a low overlap. More importantly, in-text references provide a better proxy of knowledge flow (Bryan, Ozcan, & Sampat, 2020; Marx & Fuegi, 2020a; Verberne, Chios, & Wang, 2019; Verluise, Cristelli, Higham, & de Rassenfosse, 2020). In this paper, we analyze patent in-text references for answering our research question. We also compare analytical results based on in-text and front-page references to assess whether using front-page references would lead to biased results.

The rest of this paper is organized as follows: First, we review the literature about the interaction between science and technology, focusing on how basicness, interdisciplinarity, novelty, and scientific citations may affect patent value. Second, we review the literature about in-text vs. front-page references and in particular their differences. Then we describe our data and measures. Specifically, we use the dataset developed by Voskuil and Verberne (2021) consisting of 33,337 USPTO biotech utility patents and their 860,879 references extracted from patent full texts to Web of Science (WoS) journal articles. Using this dataset, we find that (1) the number of referenced scientific papers have a positive effect on patent value, as measured by patent citations or the stock market reaction to the issuing of the patent. (2) The average basicness of referenced papers has an inverted U-shaped relation with patent value. (3) Interdisciplinarity does not have any significant effects. (4) Novelty has a discontinuous and non-linear relationship with patent value. (5) Scientific citations are positively associated with patent market value but not patent citation. We then repeat all analyses using patent front-page references instead and find that it cannot replicate results based on in-text references. To better understand these differences, we examine what types of in-text referenced papers are more likely to be listed on the front page of the same patent. We find that papers have a higher chance of being listed on front-page when they are moderately basic, less interdisciplinarity, less novel, and more highly cited. These findings contribute to the theoretical and policy debates regarding the science-technology-linkage. They also inform science and technology studies using patent references as a data source. We discuss the implications of these findings in the conclusion section.

## 2. The relationship between science and technology

Science produces fundamental understanding of natural phenomena and aims to add to the common stock of scientific knowledge, while technology is more concerned with solving real-world problems and creating practical value (Aghion, Dewatripont, & Stein, 2008; Partha & David, 1994; Stephan, 1996). Despite their fundamental differences, science and technology may feed into each other in various ways, and the complex interplay between them is a recurrent research topic (Brooks, 1994; Colyvas et al., 2002; Nelson, 1959; Tijssen, 2018).

Many scholars have argued that scientific research is valuable for technological development. Although science is less likely to yield direct practical applications, it may lead to broad, revolutionary, and unexpected applications (Bush, 1945; Partha & David, 1994; Stephan, 1996). Telling examples include: the discovery of how cells regulate the level of within-cell and external level of cholesterol (The Nobel Prize in Physiology or Medicine 1985) gave birth to the blockbuster drug, Lipitor®; and the discovery of giant magnetoresistance (The Nobel Prize in Physics 2007) has revolutionized computer hard drives. In addition to yielding direct applications, science may transform the problem-solving process underlying technological development. Fleming and Sorenson (2004) argued that science provides a map for technological development, that is, theoretical understanding of the problem and solution space can transform problem-solving from a relatively haphazard search process to a more directed identification of useful new combinations, leading to better solutions. Empirically, they found that among patents solving difficult technological problems, those citing science receive more citations from future patents than those not citing science. Others have also observed a positive association between citing science and patent value measured by patent forward citations or market value (Cassiman et al., 2008; Poege, Harhoff, Gaessler, & Baruffaldi, 2019; Watzinger & Schnitzer, 2019). In addition, Gruber, Harhoff, and Hoisl (2013) discovered that inventors with a scientific education are more likely to generate patents that span technological boundaries than inventors with an engineering degree (who are relatively more oriented toward technological development).

However, many others are skeptical about the value of science for technological development. While technological development is concerned with producing artefacts with reliable performance in various real-world situations, scientific research focuses on predictive theories and conduct research in lab settings which is an abstraction of the far more complex reality (Nelson, 2003). In addition, science and technology follow different institutional logics, e.g., motivation and incentives, performance standards, and way of organizing work (Perkmann, McKelvey, & Phillips, 2019; Sauermann & Cohen, 2010; Sauermann & Stephan, 2013). Gittelman and Kogut (2003) further contend that these differences may lead to conflicting beliefs and approaches of problem-solving, and accordingly, science has little value, and can even be counterproductive, to technological development. Empirically, they found a negative association between important scientific papers (i.e., scientific papers that are highly cited by other scientific papers) and high-impact technological inventions (i.e., patents that are highly cited by other patents). Furthermore, Ali and Gittelman (2016) found that patents by teams consisting of only MDs (who are more oriented toward technological development) are more likely to be licensed than those by teams consisting of only PhDs (who are more oriented toward scientific research) or teams with both MDs and PhDs.

In the face of these conflicting theories and empirical evidence, we lack a good understanding regarding whether science is valuable for technological development. Furthermore, we know little about what types of science are particularly useful for technological development. Science is heterogeneous, and different types of scientific research may vary in terms of how valuable they are. Examining the heterogeneity in science may provide clues for reconciling existing conflicting theories. It may also inform science policy aiming at making science more useful for economic and societal development. Despite its importance, this topic is understudied. In this paper, we investigate which kinds of science (i.e., scientific papers) will lead to more valuable technology (i.e., patents). More specifically, we focus on four characteristics of science: basicness, interdisciplinarity, novelty, and scientific citations.

### 2.1. Basicness

Whether basic or applied research is more useful for technological development, and accordingly which one should be supported more by public investment has a longstanding debate in science policy (Bush, 1945; Etzkowitz & Leydesdorff, 2000; Gibbons, 1994; Shneiderman, 2018). The spectrum from scientific research to technological development can be further refined into: basic research, applied research, and technological development. Therefore, the discussion about whether basic or applied research is more useful for technology is a natural extension of the discussion about whether science is useful or not for technology (as discussed before). As applied research is closer to technology than basic research, we might expect that applied research will be more directly useful for technology. However, prior survey research has shown that companies view scientific publications and reports (which are relatively more basic) as the most important information source from university to their R&D projects, while the importance of contract research, consulting, or patents (which are relatively more applied) are not as important (Cohen, Nelson, & Walsh, 2002; Thursby & Thursby, 2001), suggesting higher value of basic research than applied research. Ke (2020b) studied biomedical papers in the PubMed database and observed a positive association between scientific papers' degree of basicness and citations received from patents, where basicness is measured by whether a paper has cell/animal-related MeSH (more basic) terms vs. human-related MeSH terms (more applied). In this paper we will study whether building on more basic research will lead to higher patent value.

### 2.2. Interdisciplinarity

Interdisciplinary research is viewed as the solution for big scientific and societal challenges, and accordingly, it is increasingly emphasized by science policy and funding agencies. A key thesis in Gibbons (1994)'s proposal for making science more useful for the society is to move away from discipline-based mode of knowledge production to a transdisciplinary one. Through integrating perspectives from different disciplines, interdisciplinary research is expected to deliver more useful solutions for real-world problems (National Academies, 2004). On the other hand, scientific disciplines are not arbitrarily constructed but embody evolved intellectual and social systems of work organization in response to the nature of the task and the environment (Whitley, 2000). Therefore, crossing disciplinary boundaries may lead to research of low quality, due to epistemological and methodological inconstancies between the involved disciplines. It may also lead to research that is difficult to comprehend because it does not fit existing cognitive frameworks (Leahey, Beckman, & Stanko, 2016; Zuckerman, Kim, Ukanwa, & von Rittmann, 2003). Many studies have investigated whether interdisciplinary research leads to higher scientific impact and have discovered rather complex patterns (Larivière, Haustein, & Börner, 2015; Wang, Thijs, & Glänzel, 2015; Yegros-Yegros, Rafols, & D'Este, 2015). Regarding the technological impact of interdisciplinary research, Ke (2020a) studied PubMed papers and found that more interdisciplinary scientific papers are more likely to be cited by patents, where a paper's interdisciplinarity is measured by the disciplinary diversity of its references. In this paper, we will investigate whether patents building on interdisciplinary research have higher value.

### 2.3. Novelty

The value of novelty for science is self-evident, as it is novelty that pushes forward the knowledge frontier and drives scientific advancement (Merton, 1957). The usefulness of scientific novelty for technology is however less straightforward, due to its high risk/high gain nature (Foster, Rzhetsky, & Evans, 2015; Uzzi, Mukherjee, Stringer, & Jones, 2013; Wang, Veugelers, & Stephan, 2017). On the one hand, novel science might be more directly useful for technology than incremental science, as novel science is more likely to deliver breakthroughs that enable broad and revolutionary solutions. On the other hand, novel science might be less directly useful for technology because it might be premature and face a high level of uncertainty, such that series of follow-on research and development is needed before it can yield any reliable practical applications. Like interdisciplinary research, novel research may also be difficult to comprehend as it deviates from the existing paradigm. Prior studies have examined the relationship between scientific novelty and technological usefulness, and found that novel scientific papers are more likely to be cited by patents, cited in a broader set of technological fields, and lead to novel patents (Ke, 2020b; Veugelers & Wang, 2019). Veugelers and Wang (2019) studied WoS papers and measured novelty as new combinations of referenced journals, while Ke (2020b) studied PubMed papers and measured novelty as new pairs of MeSH terms. However, we still do not know whether novel science leads to more valuable patents. In addition, it is important for science policy to understand the relation between scientific novelty and technological value. Science, as a self-governed system of work organization and control, is structured to encourage novel contributions to the common stock of scientific knowledge (Merton, 1973; Stephan, 1996; Whitley, 2000). If the self-governed institutional commitment to novelty in science mismatches the policy goal of making science more useful for the society, then policy intervention is necessary for achieving the policy goal. If not, then it might be the best to leave the autonomous self-governed science system function on its own without external interference.

### 2.4. Scientific citations

Although scientific citations are noisy and subject to many sources of influence (Bornmann & Daniel, 2008; Martin & Irvine, 1983; Wang, 2014), citation-based metrics have been widely used for assessing scientific impact or quality in research evaluations. To some extent, citations have become an important currency in science. From a Mertonian perspective, citation embodies peer recognition and serves as an elementary building block of the science reward system (Merton, 1973). Therefore, examining whether science of high quality in scientific standards (or being perceived by other scientists as such) also leads to more valuable technology allows us to test whether the logic of citations in science is complementary or incompatible to that of technology. Studies have provided consistent evidence that highly cited scientific papers are more likely to be cited by patents (Ahmadpoor & Jones, 2017; Hicks, Breitzman, Hamilton, & Narin, 2000; Popp, 2017; Veugelers & Wang, 2019). However, findings are mixed regarding the association between impactful scientific papers and impactful patents (Gittelman & Kogut, 2003; Poege et al., 2019). In addition, answering this question is also important for science policy, in the same vein as for scientific novelty. The collective appropriation of task outcomes to produce new knowledge represents the second major feature of the science system in addition to the institutional commitment to novelty (Whitley, 2000). In other words, the self-governed science system builds on peer recognition and rewards research that can influence, direct, and is essential for the work of colleagues in the field. Therefore, it is important to know whether scientists' pursuit of peer recognition is in line with the policy's pursuit of practical value.

## 3. Patent in-text vs. front-page references

References in patents to scientific papers (often referred to as *scientific Non-Patent References*, *sNPRs*) provide a paper trail of the knowledge flow from science to technology. Since the pioneer work of Narin and Noma (1985), sNPRs have been widely used for studying the interaction between science and technology, for example, for investigating how companies source science for their innovative performance (Cassiman et al., 2008; Fleming & Sorenson, 2004; Marx & Fuegi, 2020b), examining concordances and distances between scientific disciplines and technological domains (Ahmadpoor & Jones, 2017; Callaert, Vervenne, et al., 2014), tracing the time lag in the knowledge diffusion process (van Raan, 2017; van Raan & Winnink, 2018), quantifying economic returns of public science (Fleming, Greene, Li, Marx, & Yao, 2019; Li, Azoulay, & Sampat, 2017; Narin, Hamilton, & Olivastro, 1997), and identifying characteristics of scientific contributions that are particularly useful for technological development (Hicks et al., 2000; Ke, 2020a, 2020b; Poege et al., 2019; Popp, 2017; Veugelers & Wang, 2019).

However, these studies only use patent *front-page references* and ignore patent *in-text references*. *Front-page references* are the references listed on the front page of the patent document, which are deemed as relevant prior art for assessing patentability. *In-text references* are references embedded in patent full text, serving a similar role as references in scientific papers. The exclusive focus on front-page references is probably due to data availability. Front-page references are easily retrievable from the meta-data of patents, although matching them to individual scientific papers is not an easy task. In comparison, extracting patent in-text references is a much more challenging task, as these references are embedded in the running text without consistent structural cues and typically contain even less information than front-page references. Thanks to the recent advancement in computer science, scholars have started developed rule-based or machine-learning methods for extracting and matching patent in-text references to scientific papers (Bryan et al., 2020; Marx & Fuegi, 2020a; Verberne et al., 2019; Voskuil & Verberne, 2021).

These recent studies have shown that in-text references embody information that is different from front-page references and provide a better proxy of knowledge flow. All these studies have observed a low overlap between front-page and in-text references, reflecting their distinct generation processes. Front-page references are generated because of patent applicants' legal duty to disclose prior art that is relevant for assessing the patentability of the focal invention. In comparison, in-text references are more like references in academic papers and may capture prior research that has enabled or influenced the focal invention but does not directly relate to patentability, for example, motivating facts, open scientific puzzles, and enabling research tools (Bryan et al., 2020; Meyer, 2000). Consistent with the process through which references are generated, several earlier studies based on surveys or interviews have suggested that front-page references may not represent a direct link between the citing patent and the cited scientific paper, but the cited scientific paper plays a more indirect role as a source of relevant background information (Callaert, Pellens, & Van Looy, 2014; Meyer, 2000; Nagaoka & Yamauchi, 2015; Tijssen, Buter, & van Leeuwen, 2000). In their work that pioneered the analysis of patent front-page references, Narin and Noma (1985) stated that patent in-text references might be a better instrument for tracing knowledge flow from science to technology. Bryan et al. (2020), who pioneered the analysis of patent in-text references, have provided further evidence supporting this statement: First, using paper-patent-pairs (which consists of a scientific paper and a patent about the same biotech research output), references in the paired paper are much more likely to be cited by the paired patent in text rather than on front page. Second, using firm survey data, the number of a company's patent in-text references to science is more strongly correlated with its reliance on science as reported by its R&D manager.

Furthermore, Marx and Fuegi (2020a) found that in-text references are older, less localized, less self-cited, more interdisciplinary, and more cited by future patents, compared with front-page references. In addition, Verluise et al. (2020) compared patent in-text references to patents with front-page references to patents. They found a low overlap between them too and that in-text referenced patents has higher text-similarity to the citing patents than front-page referenced patents. Although this study is about patent references to patents rather than to scientific papers, it is still informative for this study as it suggests that in-text referencing exhibits a stronger intellectual connection than front-page referencing, in general.

Taken together, prior studies suggest that in-text reference is a better proxy of knowledge flow and that studies using in-text reference may discover patterns different from studies using front-page references. Therefore, this paper will assess the difference in the results based on different types of references.

## 4. Data and methods

### 4.1. Data

In this paper we study how patent value is affected by referencing different types of science, in terms of basicness, interdisciplinarity, novelty, and scientific impact. We assess two types of patent references: in-text and front-page. Our paper-patent-links data come from Voskuil and Verberne (2021). They trained the state-of-the-art BERT-based machine-learning models for extracting patent in-text references to scientific papers. The pre-trained BERT models were fine-tuned on a set of 1,952 hand-labelled references in 22 patent documents. The algorithm automatically classified words into three categories: (*B*) the beginning of a reference, (*I*) inside a reference, and (*O*) outside a reference. The accuracy of the algorithm is reflected in its prediction power for *B* and *I* labels (Ramshaw & Marcus, 1999; Sang & De Meulder, 2003). The accuracy of the best-performing method, as measured in leave-one-out validation, is very high: test recall and precision are 94.7% and 95.4% respectively for beginnings of citations, and 98.6% and 97.6% for words inside citations. Subsequently, they matched the extracted in-text references (as well as front-page references) to journal articles in Web of Science (WoS) using rule-based reference parsing. The final dataset consists of all the biotech utility patents granted by USPTO from 2006 to 2010 retrieved from Google Patents, and each patent is linked to a set of its referenced WoS journal articles, in text or on front-page. In total, the dataset for our analysis consists of 33,337 patents and their 860,879 in-text and 637,570 front-page references to scientific papers in WoS.[1] Building on the paper-patent-link data, we first construct various measures for individual patents and papers. For regression analysis, our unit of analysis is a patent. We use two measures of patent value as dependent variables and construct science measures based on the profile of a patent's (in-text or front-page) scientific references as independent variables.

---

[1] The dataset from Voskuil and Verberne (2021) consists of 33,338 patents, we excluded one plant patent and kept only utility patents.

## 4.2. Patent measures (dependent variables)

For each patent, we construct two measures to capture its value: (1) *Patent citations*, which is the number of times a patent is cited by future patents, using a five-year citations time window, following the common practice. (2) *Market value*, which is based on the stock market response to the issuing of the patent, in million US dollars, developed by Kogan, Papanikolaou, Seru, and Stoffman (2017). Information about the market value is available for a subset of 7,336 patents in our sample. Both patent citations and market value have been used by Bryan et al. (2020) for comparing patent in-text and front-page references.

## 4.3. Science measures (independent variables)

We first construct measures for individual scientific papers. For *Basicness*, we adopt the measure proposed by Weber (2013) for biomedical research, which classifies a paper as *highly basic* if it only has cell/animal-related MeSH terms but no human-related MeSH terms, *moderately basic* if it has both cell/animal- and human-related MeSH terms, and *not basic* (i.e., clinical) if it only has human-related but not cell/animal-related MeSH terms. This measure is an ordinal measure, and its attributes 1, 2, and 3 correspond to *not basic*, *moderately basic*, and *highly basic*, respectively. This measure is specific for biomedical fields and is suitable for our sampled patents in the domain of biotech. This basicness measure has been used by Ke (2020b) for studying whether basic science is more cited by patents.

For *Interdisciplinarity*, we adopt the Rao-Stirling measure (Stirling, 2007), which captures all the three diversity dimensions (i.e., variety, balance, and disparity) of the involved disciplines underlying a study. More specifically, it equals $\sum_{i \neq j} p_i p_j d_{ij}$, where *i* and *j* are indices of a paper's referenced disciplines (i.e., WoS subject categories), $p_i$ is the proportion of references to discipline *i*, and $d_{ij}$ is the distance between discipline *i* and *j*, measured as 1 – cosine similarity between discipline *i* and *j* based on their co-citation matrix. This measure is a continuous measure ranging from 0 to 1. This measured has been used by Ke (2020a) for studying whether interdisciplinary papers are more cited by patents.

For *Novelty*, we adopt the measure developed by Wang et al. (2017), which follows the combinatorial novelty perspective and identifies novel paper as the ones that makes unprecedented combinations of pre-existing knowledge components, where knowledge components are proxied by referenced journals. This measure is a binary variable: 1 if novel and 0 if not novel. This measure has been used by Veugelers and Wang (2019) for studying the technological impact of novel science.

For *Scientific citations* we count the number of forward citations a scientific paper receives from future papers in the Web of Science (WoS) database, using a five-year citation time window, following the common practice. Citation counts have been widely used as the focal explanatory variable or a control variable in previous sNPR studies.

At the patent level, for quantifying a patent's profile of referenced science, in terms of basicness, interdisciplinarity, novelty, and scientific citations, we take the average of these four measures across its referenced scientific papers: *Avg*(*Basicness*), *Avg(Interdisciplinarity)*, *Avg*(*Novelty*), and *Avg*(*Scientific citations*). In addition, our focal explanatory variables also include *I*(*sNPR*), which indicates whether a patent has any scientific references, and *sNPRs*, which is the number of unique WoS papers referenced by a patent. For all these measures, we construct two versions: one based on patent in-text references and the other based on patent front-page references.

## 5. Results

### 5.1. Comparing in-text and front-page references

Before reporting regression results, we first examine the overlap between patent front-page and in-text references. The 33,337 USPTO biotech patents in our sample made 1,325,168 references to WoS papers either in text or on front page. In other words, pooling together in-text and front-page reference uncovers 1,325,168 paper-patent-links. Among them 860,879 are in-text references, and 637,570 are front-page references. Figure 1 reports the overlap between in-text and front-page references. In total, 173,281 references appear both in the text and on the front page of the same patent, which accounts for only 20% of all in-text references and 27% of all front-page references. This observed low overlap is in line with prior observations and suggests that in-text and front-page references embody different types of information (Bryan et al., 2020; Marx & Fuegi, 2020a; Verberne et al., 2019).

Insert Figure 1 here

A scientific paper can be cited by multiple patents. The 1,325,168 total references are linked to 336,522 unique papers, the 860,879 in-text references are linked to 195,988 unique papers, and the 637,570 front-page references are linked to 245,852 unique papers. Although in-text references have a larger volume (i.e., more paper-patent-links), they are linked to fewer unique papers, compared with front-page references. In other words, in-text references are concentrated in a smaller set of papers than front-page ones. In-text referenced papers are cited more often than front-page referenced papers. On average, in-text referenced papers are cited by 5.4 patents in our sample in text or on front page, 4.4 patents in text, and 1.9 patents on front-page, and the corresponding numbers are 4.6, 2.7 and 2.6 for front-page referenced papers, respectively.

We further assess the difference between in-text and front-page references in terms of their basicness, interdisciplinarity, novelty, and scientific citations. Figure 2 plots the distributions of these four measures for in-text and front-page references separately. Because the sample size is large, all the mean differences are highly significant (i.e., $p<0.001$) according to Welch two sample t-tests and Wilcoxon rank sum tests, although the difference in interdisciplinarity and novelty seem very small in size. Taken together, results show that in-text references are more basic and have more scientific citations than front-page references. In-text references are less interdisciplinary but more novel than front-page references, but the differences are small. This finding suggests that studies of which kinds of science is more cited by patents might be sensitive to whether the data come from patent in-text or front-page references.

Insert Figure 2 here

### 5.2. Patent level comparison

Descriptive statistics for patent-level variables are reported in Table 1. 80.6% of our sampled patents have in-text scientific references, while 87.3% have front-page references. Among those with in-text references, they cite on average 32.0 scientific papers in-text. Among those with front-page references, they cite on average 21.9 papers on front-page. These differences are significant according to Wilcoxon matched-pairs signed-rank test at significance level of 0.05. In the previous section, we have shown that in-text references have a larger volume but are concentrated in a smaller set of scientific papers. It appears that in-text references are also concentrated in a smaller set of patents.

Wilcoxon matched-pairs signed-rank tests also suggest that the average basicness and scientific citations of papers in a patent's in-text references are significantly higher than that of front-page references in the same patent, while there are no significant differences in average interdisciplinarity or novelty.

Correlations between the variables based on in-text and front-page references are moderate. The Spearman correlation is 0.466 between whether having in-text references and whether having front-page references. The correlations between two versions of variables (i.e., in-text and front-page) are 0.324, 0.602, 0.533, 0.261, and 0.430, for the number of referenced papers, average basicness, average interdisciplinarity, average novelty, and average scientific citations, respectively. These moderate correlations suggest that, if we rank patents by their number of scientific references or the average basicness, interdisciplinarity, novelty, and scientific citations of their referenced scientific papers, using in-text and front-page references will produce rankings that are substantially different. Furthermore, if we study the association between the characteristics of patents and the characteristics of their referenced scientific papers, we might come to different conclusions depending on whether in-text or front-page references are used.

Insert Table 1 here

### 5.3. In-text scientific references and patent value

In the next step we estimate how the characteristics of referenced science affect patent value, as measured by patent forward citations, where referenced science is based on in-text references. The dependent variable is an over-dispersed count variable, so we fit a series of Negative Binomial (NB) models. Regression results are reported in Table 2. Column 1 reports the NB model that uses whether having scientific references as the focal independent variable and incorporates the complete set of patent's issuing year and IPC class dummies. The result suggests that patents having in-text scientific references receive 29.1% more patent citations than patents not having in-text scientific references, issued in the same year and IPC class. Within the set of patents that have in-text scientific references, we further examine the intensity of reliance on science, that is, the number of referenced scientific papers. This independent variable is also a count variable and has a skewed distribution, so we take its natural logarithm for regression analysis. Column 2 shows that as a patent's number of referenced papers increases by 1%, its patent citations increase by 0.122%. These results are consistent with what Bryan et al. (2020) found.

Then we move on to explore the characteristics of referenced science. Column 3-6 each uses average basicness, interdisciplinarity, novelty, and scientific citations of referenced papers as the focal independent variable. In all these models, the *ln*(number) of scientific references is controlled for, in addition to patent issuing year and IPC class. *Avg*(*Scientific citations*) is skewed so it takes natural logarithm transformation for regression analysis. Column 3 shows that, as the average basicness of referenced papers increases by 1, patent citations decrease by 7.0%, holding all other variables fixed. Column 4 suggests no significant effects of interdisciplinarity. Column 5 shows that, as the average novelty of referenced papers increases by 1, patent citations increase by 15.6%, holding all other variables fixed. Column 6 suggests no significant effects of scientific citations. Column 7 further fits a model with all these four variables together and yields consistent results as running separate models for each independent variable (i.e., Column 3-6). In summary, patents building on less basic but more novel science are more impactful in the technological domain.

Insert Table 2 here

We then use the market value (in million US dollars) of the patent based on stock market reaction to the event of patent being issued as the dependent variable. This variable is also skewed but not a count variable, so we take natural logarithm transformation and then fit Ordinary Least Squares (OLS) models. Results are reported in Table 3. Results show that patents with in-text scientific references worth 86.8% more than patent without in-text scientific references (Column 1). Within the set of patents having in-text scientific references, patent market value increases by 0.190% as the number of referenced scientific papers increases by 1% (Column 2). As the average basicness of referenced science increases by 1, patent market value decreases by 37.8% (Column 3). Interdisciplinarity and novelty has no significant effects on patent market value (Column 4 and 5). As the average scientific citations of referenced science increase by 1%, patent market value increases by 0.094% (Column 6). These results are robust when fitting a model with all four variables together (Column 7). In summary, patents building on papers that are less basic but more highly cited in science generate higher private market value.

Insert Table 3 here

### 5.4. Complexity in the effects

Our regression models assume a linear equation, where the left-hand side is the natural log of one dependent variable (patent citations or market value), and the right-hand side consists of a series of independent variables. This setup is flexible for fitting positive (or negative) effects at an increasing or decreasing rate. However, it does not allow nonmonotonic effects (e.g., inverted U-shaped) or discontinuous effects. Therefore, we reexamine the effects using a more non-parametric approach without assuming a linear equation. Specifically, we categorize our independent variables into 10 ordered and evenly sized groups and then estimate the expected patent citations and market value for each group. Because of ties, not all groups are evenly sized. Taking the number of in-text scientific references as an example, 6,465 patents with 0 references are classified into Group 1, the next 1,891 patents with only 1 reference are classified into Group 2, …, the last 3,330 patents with 71 to 711 references are classified into Group 10. Group assignments for all the independent variables are reported in Appendix I Table A1. Then we use group numbers as a factor/categorical variable for regression analysis, and results are reported in Table 4 and 5. These regressions essentially estimate the differences between the reference group (i.e., Group 1, omitted in the regression table) and the other nine groups (Group 2-9 reported in the regression table), controlling for other control variables. Based on the regression results, we can estimate the expected value of the dependent variables for each group for an average patent (i.e., issuing year is 2010, IPC class is C12N, and other control variables (if any) at the mean). Figure 3 plots these estimates.

Insert Table 4 here

Insert Table 5 here

Insert Figure 3 here

Regarding the number of in-text references, consistent with the result reported in the preceding section, Fig 3A&F display a roughly continual increase in patent citations and market value, as the number of in-text references move from 0 (Group 1) to 1 (Group 2), and then further increases (from Group 2 to Group 10). There is consistent evidence that citing science and citing more scientific papers have a positive effect on patent value.

For average basicness, the previously reported result suggests that it has negative effects on patent citations and market value. However, results in Fig 3B&G suggest inverted U-shaped effects. Both patent citations and market value first increase and then decrease as the average basicness increases. Patent citations reach the peak point at Group 2, and if we dismiss Group 2 due to its small group size, then the peak point is reached at Group 4. Patent market value reaches its peak at Group 4. These results suggest that a moderate level of basicness is optimal for patent value, while too applied or too basic lead to lower patent value.

In terms of interdisciplinarity, our previous result suggests insignificant effects of interdisciplinarity on patent citations and market value. Fig 3C seems to suggest a U-shaped effect, but the p-value for the joint significance of interdisciplinarity groups is larger than 0.05. Fig 3H suggests no clear association between average interdisciplinarity and patent market value. Taken together, we conclude no significant effect of average interdisciplinarity on patent value.

Our previous result suggests that average novelty has a positive effect on patent citations but an insignificant effect on patent market value. Fig 3D&I reveal more complex patterns. According to Fig 3D, as a patent moves from having no novel references to having novel references, there is a sudden drop in patent citations. As the average novelty further increases, patent citations rise and slowly reach a plateau (or even go down). According to Fig 3I, there is a disruptive rise in patent market value when moving from having no novel references to having novel references. However, as the average novelty further increases, patent market value decreases and slowly flattens (or even bounces up). Taken together, these results suggest a structural change between patents building on novel science and those not. As a patent builds on novel science, its technological impact drops, potentially due to uncertainties introduced by sourcing novel science. Its technological impact then recovers and reaches a higher point than patents not building on novel science, indicating that sourcing more novel science leads to broader and more unexpected applications. However, this increasing trend does not continue unlimitedly, the benefit from sourcing novel science stops at certain level of average novelty. Regarding patent market value, sourcing novel science brings a jump in the stock market reaction to the patented technology, reflecting market's appreciation of novelty. However, further increase in the novelty of sourced science reduces market value as the patent might become too remote from marketable applications.

Consistent with our previous result that average scientific citations have an insignificant effect on patent citations but a positive effect on patent market value, Fig 3E exhibits no clear associations between scientific citations and patent citations, but only fluctuations around a flat line, and Fig 3J displays an increasing trend with some fluctuations. Therefore, we conclude an insignificant effect of scientific citations on patent citations but a positive effect on patent market value. This suggests that the criteria of usefulness might not be perfectly aligned between science and technology, scientific outputs that are (perceived) useful for others to do follow-on scientific research (i.e., receive more scientific citations) do not necessarily leads to technologies that are (perceived) useful for others to develop follow-on technologies (i.e., receive more patent citations). On the other hand, there is neither a constriction between them as no significantly negative relation is observed. Regarding patent market value, scientific outputs that are highly recognized by other scientists are positively associated with technologies that are highly appreciated by the stock market, reflecting a certain level of alignment between scientists' interest and the market's interest.

### 5.5. Front-page scientific references and patent value

We repeat all the analyses using front-page references instead, to test whether using front-page will lead to the same findings. Figure 4 repeats Figure 3 but additionally overlay it with estimates based on patent front-page references. Regression outputs replicating Table 2-5 are reported in Appendix II Table A2-A5.

Insert Figure 4 here

Results on the number of front-page scientific references are consistent without what is observed when using in-text scientific references (Fig 4A&F), which display a continually positive effect of the number of scientific references on patent value.

Regarding average basicness, results are substantially different. While the inverted U-shaped effect of average basicness on patent market value with an overall negative trend remains (Fig 4G, Table A3 Column 3), there are no clear associations between average basicness and patent citations (Fig 4B, Table A2 Column 3).

While the insignificant effect of average interdisciplinarity on patent market value remains (Fig 4H, Table A3 Column 4), there is a significantly positive effect of average interdisciplinarity on patent citations (Fig 4C, Table A2 Column 4).

No clear patterns or significant associations between average novelty and patent value are displayed (Fig 4D&I, Table A2&3 Column 5). Fig4D further shows that the general trend based on front-page references still resembles that of in-text references, but the size of effect is much smaller and become insignificant.

While the positive effect of average scientific citations on patent market value remains (Fig 4J, Table A3 Column 6), the effect of average scientific citations on patent citations become weakly significantly positive (Fig 4E, Table A2 Column 6).

In summary, using front-page reference cannot replicate the results based on in-text references. More specifically, when examining patent market value, results based on front-page references are largely consistent with results based on in-text references, except that the effect of novelty becomes insignificant. When examining patent citations, results are substantially different: Effects of basicness and novelty become insignificant, while insignificant effects of interdisciplinarity and scientific citations become significantly positive.

### 5.6. Why do front-page and in-text references yield different results?

The inconsistencies between the results based on front-page and in-text references are not surprising, considering their low overlap and the moderate correlations between science measures based on front-page and in-text references. In this section we attempt to explore why such inconsistences emerge, by looking into the processes through which in-text and front-page references are generated. As discussed before, in-text references document various sources of knowledge that are instrumental to the patented technology, while front-page references are listed for disclosing prior arts that are relevant for assessing patentability. Sampat (2010) argued that, for patents that are expected to be more valuable, patent applicants may perform a more comprehensive prior art search, to prevent the chance that the patent application is rejected due to failure of disclosure. This more comprehensive search may result in a longer list of front-page references. Therefore, we view in-text references as an unbiased (but noisy) representation of the scientific outputs underlying a focal patented technology. In comparison, front-page references also reflect this unbiased representation but are subject to additional biases introduced by patent applicants' strategic behavior. It is possible that certain types of scientific papers are valuable for inspiring the patented technology but not so relevant for assessing its patentability and therefore are not listed on the front page. To explore this, we analyze individual in-text references and examine which types of in-text referenced papers are more likely to be listed on the front page of the same patent.

Insert Table 6 here

Using in-text references (i.e., paper-patent-links) as the unit of analysis, we fit conditional fixed-effects logistic models, with patent fixed effects to account for patent heterogeneities. Regression results are reported in Table 6. Column 1 shows that, for the same patent, among its in-text referenced papers, moderately basic papers have the highest chance of being listed on its front page, followed by highly basic, and lastly not basic papers. Column 2 and 3 show that more interdisciplinary and novel papers among the in-text referenced papers of the same patent are less likely to be listed on the front page of that patent. In contrast, papers with more scientific citations have a higher chance of being listed on the front page of the same patent (Column 4).

The inverted U-shaped relationship between basicness and the likelihood of being listed on patent front page is in line with the observed inverted U-shaped relationship between average basicness and patent value. This means that a moderate level of basicness is not only positively associated with higher patent value but also higher degree of relevance for assessing patentability. Both interdisciplinarity and novelty deviates from the existing paradigm, and their contribution to the patented technology might be rather unexpected. Therefore, their intellectual link to the patent is relatively distant and tenuous, and their relevance for assessing patentability is relatively low. On the other hand, highly cited papers have generated more follow-on research and therefore is also possible to have more direct relevance for assessing patentability. In addition, highly cited papers are more visible in both domains of science and technology, such that missing them would bring a higher risk of being rejected due to failure of disclosure. Since front-page references systematically under-represent interdisciplinary and novel papers but over-represent moderately basic and highly cited papers. We can expect that using front-page references will yield substantially different results than using in-text references when analyzing these science measures.

### 5.7. Robustness tests

The average basicness, interdisciplinarity, novelty, and scientific citations measures might have more fluctuations and be less reliable when the number of references is small.

Therefore, we test the robustness of our results by repeating the analyses using patents with at least 10 references. Results are consistent (Appendix III Figure A1).

We run another robustness test by repeating the analysis using patents with at least one applicant and at least one inventor located in the United States, as the referencing behavior might be different across countries. Results based on the US subsample closely resembles results based on the full sample (Appendix III Figure A2). One noticeable change is that the number of in-text references have a much smaller effect on patent market value.

**6. Discussion**

This study has several limitations, calling for further research. First, our dataset covers a sample of patents and all their references. However, we do not know whether the papers in our dataset are cited by patents outside our sample. This data limitation does not allow us to use scientific papers as the unit of analysis to investigate how different types of science are cited by all patents and how their technological impact unfolds. Second, we adopt a nonparametric approach (i.e., categorizing patents into groups based on certain science measures) to uncover the complex associations between the characteristics of referenced science and patent value. This approach is simple but powerful for this exploratory research aiming at uncovering interesting patterns. Future research should develop more formal models to systematically explain these patterns. In particular, the complex association between novelty and patent value need to be further investigated. Future research also needs to develop more sophisticated designs to allow causal inference. Third, results for patent citations are not identical to the result for patent market value. We need more research into the difference between these two measures, as well as their different response to the characteristics of referenced science. Fourth, the market value information is only available for patents from publicly listed companies, and therefore the findings about market value might not be generatable to other patents. Fifth, our analysis of the process through which in-text and front-page references are generated is very preliminary. To fully understand why front-page and in-text references lead to different analytical results, we need future research to further investigate the reference generation process, by collecting survey, interview, and archival data, and to design more rigorous and direct empirical tests. Sixth, our sample only covers USPTO biotech patents. Our findings might not be generalizable to other technological fields. How science and technology interact with each other may be different across fields. Our basicness measure is specific to the biomedical sciences but may not apply to other fields, such that the observed effects regarding basicness can only be interpreted in the context of biotech research and development. In addition, our findings may not apply to other patent authorities. Considering that the practice at USPTO is very different from that at EPO, as well as other patent offices, it is important to understand how the reference generating process might differ across patent authorities and how these differences might influence the validity of using patent references as a proxy of knowledge flow.

## 7. Conclusion

This paper investigated how patent value depends on the characteristics of scientific papers underlying the patented technology, specifically, basicness, interdisciplinarity, novelty, and scientific citations. Using a dataset consisting of 33,337 USPTO biotech patents and their 860,879 references extracted from patent full texts to Web of Science journal articles, we found that patents that cite more scientific papers are more valuable than patents that cite fewer scientific papers, where patent value is measured by the number of citations a patent receives from future patents and the market value based on the stock market response to the issuing of the patent. Within the set of patents that cite science, the average basicness of referenced science has an inverted U-shaped relation with patent value. Interdisciplinarity does not have any significant effects on patent citations or market value. The average scientific citations have an insignificant effect on patent citations but a positive effect on patent market value. Effects of novelty are complex. Patent citations have a sudden drop when moving from not citing any novel papers to citing novel papers, and then slowly rise to a higher level. On the other hand, patent market value has a sudden jump when moving from not citing any novel papers to citing novel papers, and then slowly declines.

In addition, this paper examined differences between patent in-text and front-page references. Using front-page references cannot replicate the results based on in-text references regarding the relation between the characteristics of referenced science and patent value. Results are substantially different, especially for patent citations. We further examined what types of in-text referenced papers are more likely to be listed on the front page of the same patent. We found that papers are more likely to be listed on the front page when they are moderately basic, less interdisciplinary, less novel, and more highly cited in science.

This study contributes to the recurrent debate regarding the relationship between science and technology, more specifically, whether scientific research is useful for, or rather incompatible with, technological development. Our results suggest that both sides have merits and that unpacking the heterogeneity in scientific outputs provides a promising direction to reconcile these competing theories. Our results show that patent value response to basicness, interdisciplinarity, novelty, and scientific citations in different ways. Our results also contribute to the studies of patent references to the scientific literature. The inconsistencies between the results based on in-text and front-page references indicate that studies using front-page references might lead to biased results due to its special process of generating references. However, what exactly theses biases might be, and how exactly they will influence analytical results still need further research.

Our findings also have implications for science policy. There is an increasing policy push to make science, in particular public funded scientific research in universities and public research organizations, more directly and immediately useful for economic and societal development. Our results provide clues for what kinds of science should be promoted for achieving this policy goal. More specifically, it partly supports the policy agenda advocated by Mode 2 for supporting more application-oriented research, as patent value peaks at a relatively low level of basicness. On the other hand, it warns that completely dismissing basic research is detrimental as the association between basicness and patent value is not a simple negative relation. Our results do not provide evidence that interdisciplinary research is the key for making science more useful for technological development. With respect to novelty, our results do not provide a clear message whether science policy should support novel or non-novel research, as the association between novelty and patent value is rather complex. Our results do suggest that novelty plays a special role for technological development and that innovators should be careful about potential disruptions and uncertainties that sourcing novel science brings. In terms of scientific citations, we find a positive effect of scientific citations on patent market value but not on patent citations. However, we do not see any negative effects. Therefore, we still observe a certain degree of alignment between the policy's pursuit of practical value and science's institutional committee to produce knowledge that is useful for peers' research. We do not see that letting science pursue peer recognition would harm its practical value. Therefore, the autonomous work organization of science based on reputation and peer recognition should be protected.

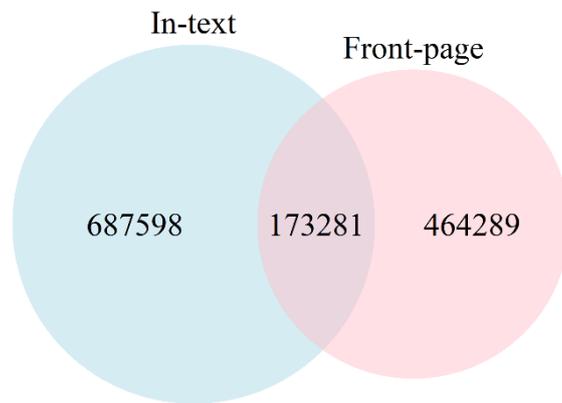

**Figure 1. Overlap between in-text and front-page references.**

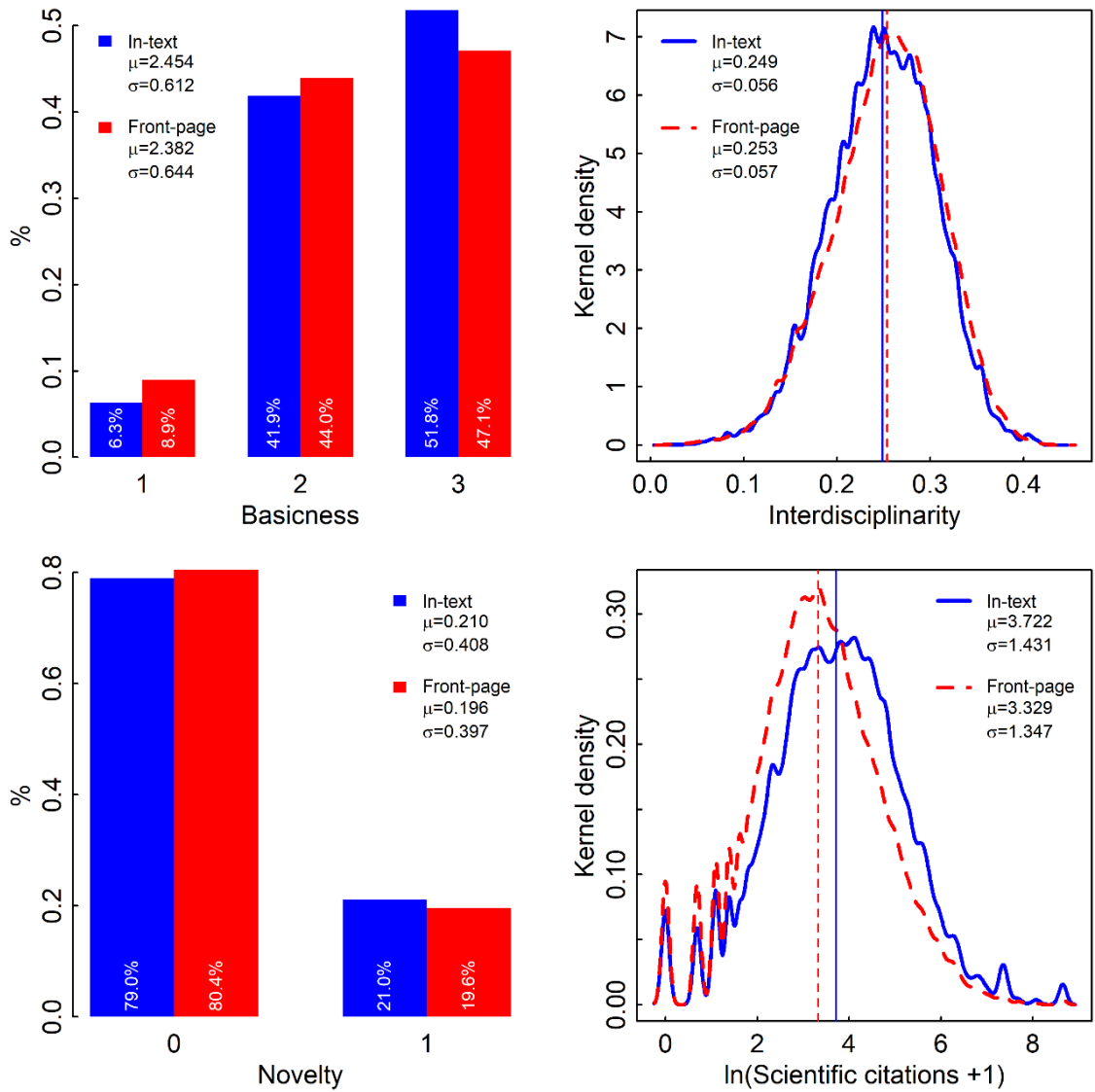

**Figure 2. Distribution of basicness, interdisciplinarity, novelty, and scientific citations, by in-text and front-page references.** Plots for basicness and novelty are simple proportions by category. Plots for interdisciplinarity and *ln*(Scientific citations +1) are kernel densities where the vertical lines mark the mean values. Scientific citations are logarithm transformed because it is highly skewed.

**Table 1. Descriptive statistics** (Unit of analysis: patent)

|  | N | Mean | Std. Dev. | Min | Max |
|---|---|---|---|---|---|
| Patent citations | 33,337 | 3.766 | 8.238 | 0 | 549 |
| Market value (m$) | 7,336 | 46.757 | 87.453 | 0.001 | 993.601 |
| *In-text* | | | | | |
| I(sNPR) | 33,337 | 0.806 | 0.395 | 0 | 1 |
| sNPRs | 26,872 | 32.036 | 43.005 | 1 | 711 |
| Avg(Basicness) | 26,012 | 2.445 | 0.394 | 1 | 3 |
| Avg(Interdisciplinarity) | 26,709 | 0.254 | 0.032 | 0.037 | 0.429 |
| Avg(Novelty) | 26,872 | 0.155 | 0.157 | 0 | 1 |
| Avg(Scientific citations) | 26,872 | 124.208 | 191.441 | 0 | 5871 |
| *Front-page* | | | | | |
| I(sNPR) | 33,337 | 0.873 | 0.333 | 0 | 1 |
| sNPRs | 29,110 | 21.902 | 31.368 | 1 | 1064 |
| Avg(Basicness) | 27,999 | 2.401 | 0.435 | 1 | 3 |
| Avg(Interdisciplinarity) | 28,982 | 0.256 | 0.037 | 0.023 | 0.419 |
| Avg(Novelty) | 29,110 | 0.162 | 0.178 | 0 | 1 |
| Avg(Scientific citations) | 29,110 | 62.580 | 88.399 | 0 | 5871 |

**Table 2. In-text scientific references and patent citations**

| | Patent citations NB | | | | | | |
|---|---|---|---|---|---|---|---|
| | (1) | (2) | (3) | (4) | (5) | (6) | (7) |
| I(sNPR) | 0.291*** | | | | | | |
| | (0.029) | | | | | | |
| ln(sNPRs) | | 0.122*** | 0.133*** | 0.122*** | 0.122*** | 0.124*** | 0.136*** |
| | | (0.011) | (0.012) | (0.011) | (0.011) | (0.011) | (0.012) |
| Avg(Basicness) | | | -0.070* | | | | -0.073* |
| | | | (0.034) | | | | (0.035) |
| Avg(Interdisciplinarity) | | | | 0.622 | | | 0.219 |
| | | | | (0.446) | | | (0.506) |
| Avg(Novelty) | | | | | 0.156+ | | 0.175* |
| | | | | | (0.082) | | (0.086) |
| ln(Avg(Scientific citations) +1) | | | | | | -0.009 | -0.009 |
| | | | | | | (0.013) | (0.014) |
| Issue year | Y | Y | Y | Y | Y | Y | Y |
| IPC class | Y | Y | Y | Y | Y | Y | Y |
| N | 33337 | 26872 | 26012 | 26709 | 26872 | 26872 | 25930 |
| BIC | 152066 | 123120 | 118841 | 122524 | 123115 | 123130 | 118555 |

Unit of analysis: patent. The dependent variable is the number of patent forward citations. Each column reports one Negative Binomial regression model. Column 1 estimates the effect of having scientific references. Within the set of patents citing scientific papers, Column 2 estimates the effect of the natural log number of referenced papers. Column 3-6 estimate effects of the average basicness, interdisciplinarity, novelty, and scientific citations separately, controlling for the number of referenced papers. Column 7 fits a model with all these five variables together. All models control for the complete set of patent issuing year and IPC class dummies. Robust standard errors in parentheses. *** $p < .001$, **$p < .01$, *$p < .05$, +$p < .10$.

**Table 3. In-text scientific references and patent market value**

| | ln(Market value) | | | | | | |
|---|---|---|---|---|---|---|---|
| | OLS | | | | | | |
| | (1) | (2) | (3) | (4) | (5) | (6) | (7) |
| I(sNPR) | 0.868*** | | | | | | |
| | (0.079) | | | | | | |
| ln(sNPRs) | | 0.190*** | 0.191*** | 0.182*** | 0.190*** | 0.164*** | 0.163*** |
| | | (0.019) | (0.020) | (0.020) | (0.019) | (0.020) | (0.021) |
| Avg(Basicness) | | | -0.378*** | | | | -0.380*** |
| | | | (0.081) | | | | (0.082) |
| Avg(Interdisciplinarity) | | | | -0.523 | | | 0.122 |
| | | | | (0.929) | | | (1.095) |
| Avg(Novelty) | | | | | 0.003 | | -0.073 |
| | | | | | (0.189) | | (0.223) |
| ln(Avg(Scientific citations)+1) | | | | | | 0.094** | 0.116*** |
| | | | | | | (0.029) | (0.032) |
| Issue year | Y | Y | Y | Y | Y | Y | Y |
| IPC class | Y | Y | Y | Y | Y | Y | Y |
| N | 7336 | 6181 | 5984 | 6143 | 6181 | 6181 | 5969 |
| R2 | 0.091 | 0.074 | 0.077 | 0.073 | 0.074 | 0.076 | 0.079 |
| BIC | 30944 | 25341 | 24427 | 25197 | 25341 | 25337 | 24374 |

Unit of analysis: patent. The dependent variable is the natural log of the market value. Each column reports one OLS regression model. Column 1 estimates the effect of having scientific references. Within the set of patents citing scientific papers, Column 2 estimates the effect of the natural log number of referenced papers. Column 3-6 estimate effects of the average basicness, interdisciplinarity, novelty, and scientific citations separately, controlling for the number of referenced papers. Column 7 fits a model with all these five variables together. All models control for the complete set of patent issuing year and IPC class dummies. Robust standard errors in parentheses. *** $p < .001$, **$p < .01$, *$p < .05$, +$p < .10$.

**Table 4. In-text scientific references and patent citations: A nonparametric approach**

| | Patent citations NB | | | | |
|---|---|---|---|---|---|
| | (1) | (2) | (3) | (4) | (5) |
| Group: | sNPRs | Basicness | Interdisciplinarity | Novelty | Scientific citations |
| 2 | 0.018 | 0.170** | -0.063 | / | 0.009 |
| | (0.048) | (0.061) | (0.058) | | (0.053) |
| 3 | 0.153* | 0.027 | -0.069 | -0.176** | 0.105 |
| | (0.074) | (0.051) | (0.060) | (0.056) | (0.077) |
| 4 | 0.040 | 0.066 | -0.056 | -0.094 | -0.018 |
| | (0.043) | (0.054) | (0.062) | (0.058) | (0.054) |
| 5 | 0.171*** | -0.062 | -0.066 | -0.039 | 0.099+ |
| | (0.041) | (0.054) | (0.060) | (0.054) | (0.057) |
| 6 | 0.401*** | -0.018 | -0.138* | 0.026 | -0.012 |
| | (0.046) | (0.049) | (0.059) | (0.056) | (0.058) |
| 7 | 0.352*** | -0.048 | -0.009 | 0.047 | 0.012 |
| | (0.045) | (0.051) | (0.059) | (0.054) | (0.060) |
| 8 | 0.275*** | -0.067 | 0.003 | 0.040 | -0.226*** |
| | (0.043) | (0.051) | (0.057) | (0.051) | (0.062) |
| 9 | 0.432*** | -0.027 | -0.022 | 0.161** | -0.030 |
| | (0.042) | (0.045) | (0.057) | (0.050) | (0.061) |
| 10 | 0.616*** | / | 0.090 | 0.021 | 0.113+ |
| | (0.044) | | (0.077) | (0.070) | (0.061) |
| ln(sNPRs) | / | 0.129*** | 0.135*** | 0.127*** | 0.130*** |
| | | (0.013) | (0.010) | (0.013) | (0.012) |
| Issue year | Y | Y | Y | Y | Y |
| IPC class | Y | Y | Y | Y | Y |
| N | 33337 | 26012 | 26709 | 26872 | 26872 |
| BIC | 151854 | 118890 | 122573 | 123135 | 123129 |
| Chi2 | 314*** | 22** | 16+ | 44*** | 53*** |
| LR Chi2 | 54*** | 22** | 22** | 51*** | 73*** |

This table repeats the analysis reported in Table 2 but uses the categorized science measures as independent variables instead. Take the variable *sNPRs Group* as an example, we code it as 1 if a patent's number of referenced papers is among the lowest 10%, 2 if among the next 10%, … 10 if among the top 10%. The complete categorization scheme is reported in Appendix Table A1. Due to ties, not all groups are evenly sized, and sometimes two groups are be merged. For example, for Avg(Basicness), group 9 and 10 are merged and labelled as group 9; for Avg(Novelty), group 1 and 2 are merged and labelled as group 1. Then we use these categorical variables as the independent variable for regression. Group 1 is the reference group. Coefficients indicates the difference between a focal group and the reference group (i.e., Group 1). Take Column 1 as an example, the coefficient of 0.018 for Group 2 means that patents in Group 2 have 1.8% more patent citations than Group 1. Chi2 tests the joint significance of all the levels of the group variable (H0: all coefficients of Group 2, 3, …,10 equal to 0). LR Chi2 reports the likelihood ratio test between the focal regression model and the model with the raw uncategorized independent variable. More specifically, it tests the model in Table 4 column 2 against the model in Table 2 column 3, Table 4 Column 3 vs. Table 2 Column 4, Table 4 Column 4 vs. Table 2 Column 5, and Table 4 Column 5 vs. Table 2 Column 6. For Table 4 Column 1, we fit another model with *ln*(sNPRs +1) as the independent variable, because Table 2 Column 2 only includes patents with at least one reference. If we test Table 4 Column 1 against Table 2 Column 1, then the result is: 295***. Being significant here means the model in this table fits the data better than its corresponding model in Table 2. Robust standard errors in parentheses. *** p < .001, **p < .01, *p < .05, +p < .10.

**Table 5. In-text scientific references and patent market value: A nonparametric approach**

| | ln(Market value) OLS | | | | |
|---|---|---|---|---|---|
| | (1) | (2) | (3) | (4) | (5) |
| Group | sNPRs | Basicness | Interdisciplinarity | Novelty | Scientific citations |
| 2 | 0.070 | 0.298* | 0.470*** | / | 0.234+ |
| | (0.146) | (0.139) | (0.118) | | (0.139) |
| 3 | 0.447*** | 0.199+ | 0.381** | 0.495*** | -0.010 |
| | (0.122) | (0.116) | (0.115) | (0.112) | (0.135) |
| 4 | 0.662*** | 0.502*** | 0.412*** | 0.737*** | -0.025 |
| | (0.120) | (0.115) | (0.115) | (0.107) | (0.137) |
| 5 | 0.835*** | 0.013 | 0.107 | 0.333** | 0.015 |
| | (0.107) | (0.115) | (0.116) | (0.112) | (0.135) |
| 6 | 0.955*** | -0.428*** | 0.225+ | 0.213+ | 0.066 |
| | (0.101) | (0.116) | (0.117) | (0.112) | (0.134) |
| 7 | 0.979*** | -0.105 | 0.356** | 0.075 | 0.079 |
| | (0.100) | (0.108) | (0.118) | (0.111) | (0.134) |
| 8 | 1.042*** | -0.346** | 0.439*** | 0.146 | 0.615*** |
| | (0.099) | (0.116) | (0.125) | (0.108) | (0.133) |
| 9 | 1.055*** | -0.181+ | 0.052 | 0.054 | 0.257+ |
| | (0.100) | (0.107) | (0.127) | (0.108) | (0.133) |
| 10 | 1.193*** | / | 0.251+ | 0.262* | 0.337* |
| | (0.092) | | (0.132) | (0.114) | (0.131) |
| ln(sNPRs) | / | 0.168*** | 0.166*** | 0.126*** | 0.163*** |
| | | (0.022) | (0.022) | (0.026) | (0.022) |
| Issue year | Y | Y | Y | Y | Y |
| IPC class | Y | Y | Y | Y | Y |
| N | 7336 | 5984 | 6143 | 6181 | 6181 |
| R2 | 0.107 | 0.093 | 0.079 | 0.087 | 0.086 |
| BIC | 30865 | 24374 | 25206 | 25314 | 25334 |
| F | 26*** | 16*** | 5*** | 12*** | 9*** |
| LR Chi2 | 32*** | 106*** | 43*** | 88*** | 64*** |

This table repeats the analysis reported in Table 3 but uses the categorized science measures as independent variables instead. F tests the joint significance of all the levels of the group variable (H0: all coefficients of Group 2, 3, …,10 equal to 0). LR Chi2 reports the likelihood ratio test between the focal regression model and the model with the raw uncategorized independent variable. More specifically, it tests the model in Table 5 column 2 against the model in Table 3 column 3, Table 5 Column 3 vs. Table 3 Column 4, Table 5 Column 4 vs. Table 3 Column 5, and Table 5 Column 5 vs. Table 3 Column 6. For Table 5 Column 1, we fit another model with *ln*(sNPRs +1) as the independent variable, because Table 3 Column 2 only includes patents with at least one reference. If we test Table 5 Column 1 against Table 3 Column 1, then the result is: 133***. Being significant here means the model in this table fits the data better than its corresponding model in Table 2. Robust standard errors in parentheses. *** p < .001, **p < .01, *p < .05, +p < .10.

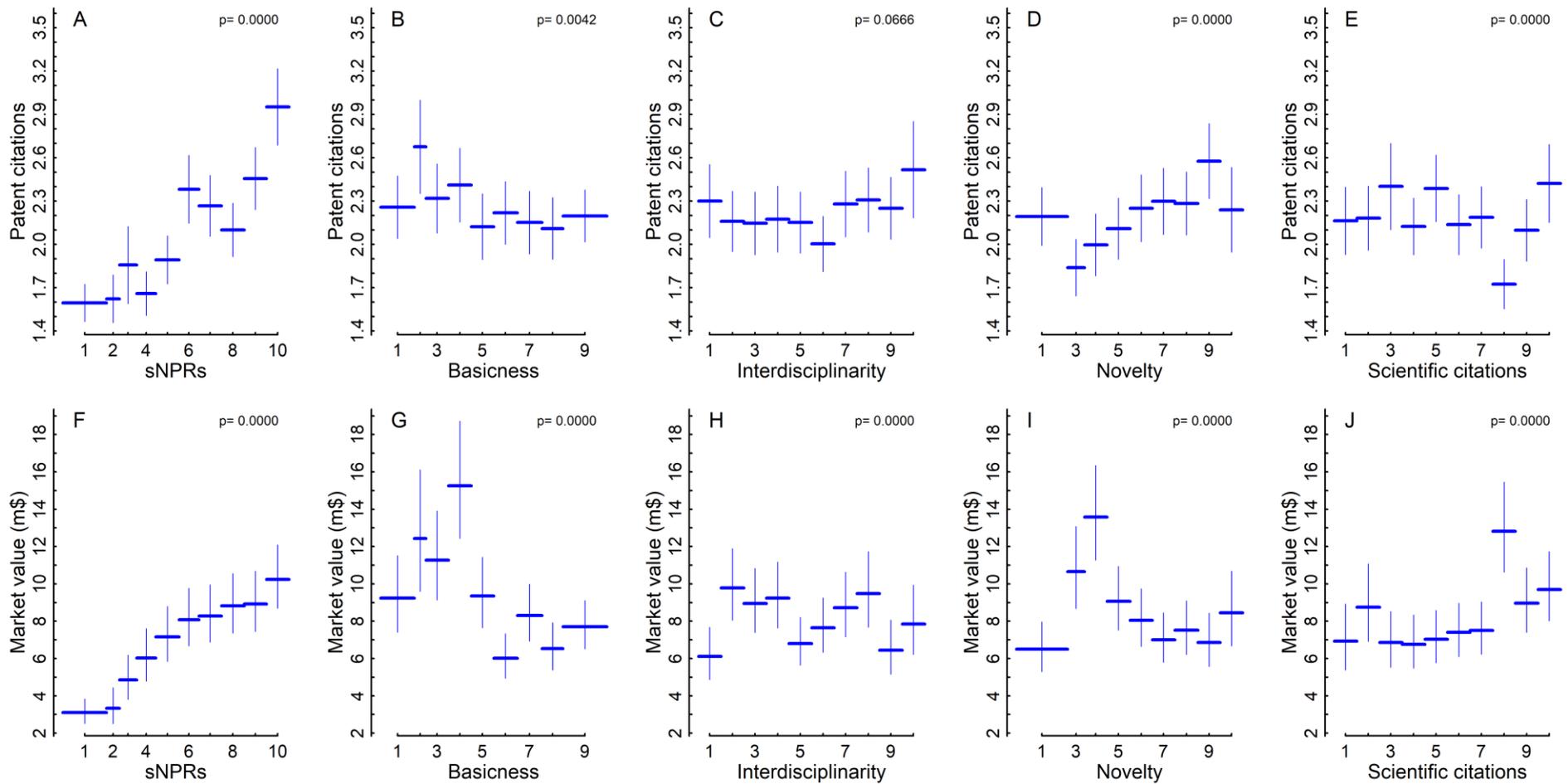

**Figure 3. In-text scientific references and patent value**. This figure plots the estimated value of patent value for an average patent in different science measure groups. Plot A-E and F-J are based on regression models reported in Table 4 Column 1-5 and Table 5 Column 1-5, respectively. An average patent means its issuing year is 2010, IPC is C12N, and for Plot B-E and G-J, the natural log number of references takes the mean value. The length of the blue horizontal lines is proportional to the group size. The blue vertical lines mark the 95% confidence interval. p-value is for the joint significance of all the levels of the group variable, corresponding to Chi2 in Table 4 and F in Table 5.

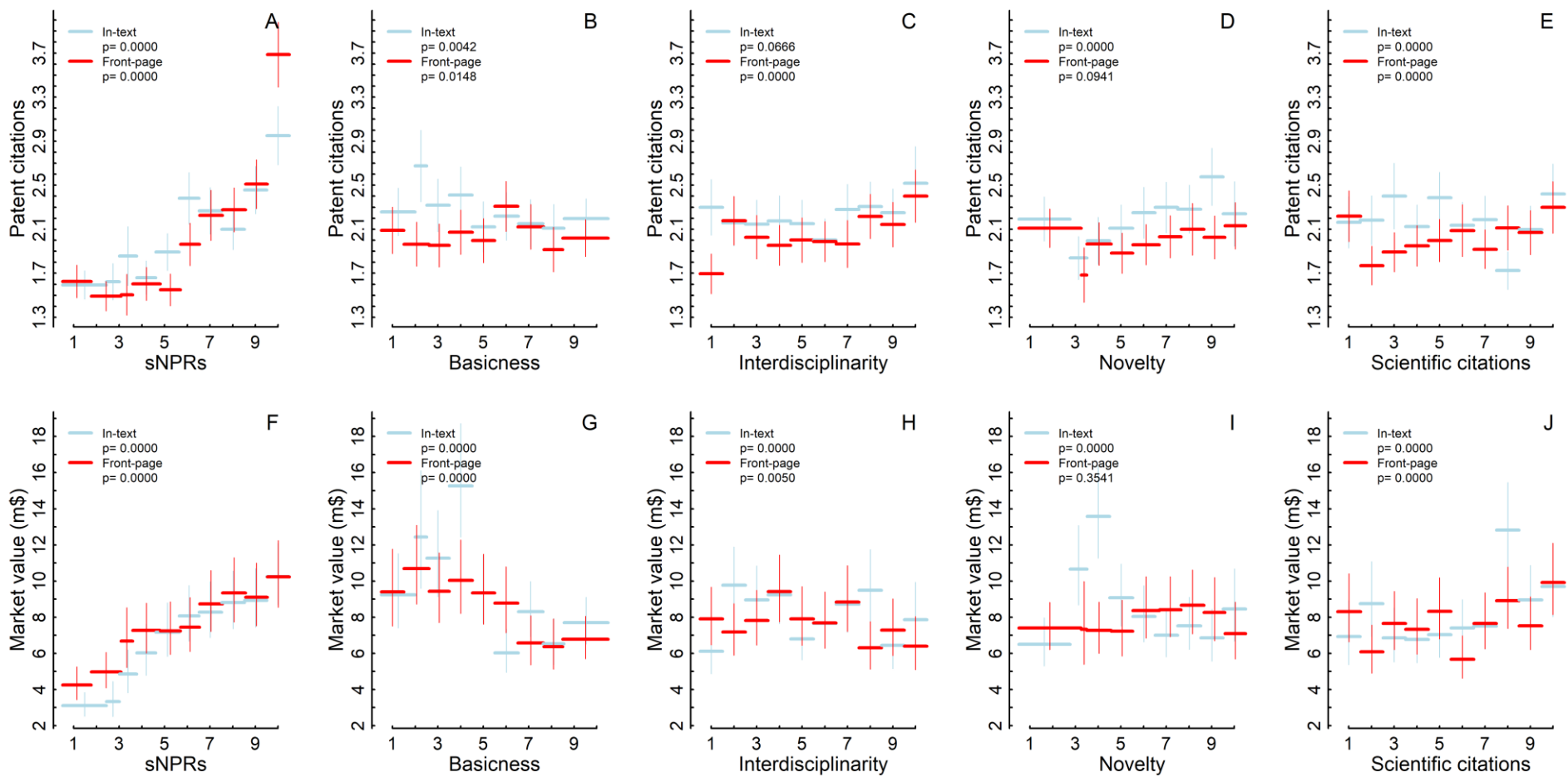

**Figure 4. Scientific references and patent value: Front-page vs. in-text**. The light blue lines in this plot are identical to the blue lines in Figure 3. They represent results based on in-text references. We overlay them with the red lines (results based on front-page references) for an easier comparison. Specifically, we repeat the same procedure for producing Figure 3 but use science measures based on patent front-page references instead. Patent value estimates for the red lines are based on regression results reported in Table A4 and Table A5.

**Table 6. What types of in-text referenced papers are more likely to be listed on the patent front page?**

|  | I(Front-page) | | | |
|---|---|---|---|---|
|  | Conditional fixed-effects logit | | | |
|  | (1) | (2) | (3) | (4) |
| Basicness=1 | -0.225*** | | | |
|  | (0.019) | | | |
| Basicness=2 | 0.106*** | | | |
|  | (0.009) | | | |
| Interdisciplinarity | | -0.638*** | | |
|  | | (0.077) | | |
| Novelty | | | -0.025* | |
|  | | | (0.011) | |
| ln(Scientific citations +1) | | | | 0.065*** |
|  | | | | (0.003) |
| Paper: Publication year | Y | Y | Y | Y |
| Paper: Scientific field | Y | Y | Y | Y |
| N obs | 570408 | 634225 | 613065 | 655311 |
| N patents | 15374 | 16217 | 16093 | 16394 |

Unit of analysis: in-text references, i.e., paper-patent-links through in-text referencing. All models incorporate patent fixed effects, so that estimates are about within-patent differences. The dependent variable I(Front-page) is a binary variable: 1 if an in-text referenced paper is listed on the front page of the same patent, and 0 otherwise. Each column reports one conditional fixed-effects logistic regression model. For Column 1 we treat basicness as a categorical variable with three levels: 1, 2, and 3. Level 3 is used as the reference group, so the coefficient reports the difference between a focal basicness group and the reference group. For example, the coefficient of -0.225 means the log-odds of being listed on the front-page (vs. not being listed) is 0.225 smaller for patents that are not basic (Basicness =1) than for patents that are highly basic (Basicness = 3, the reference group). Robust standard errors in parentheses. *** $p < .001$, **$p < .01$, *$p < .05$, +$p < .10$.

**Appendix I. Categorizing science measures based on in-text references.**

**Table A1. Scheme for categorizing science measures based on in-text references into 10 groups.**

| | Min | Max | N |
|---|---|---|---|
| sNPRs Group | | | |
| 1 | 0 | 0 | 6,465 |
| 2 | 1 | 1 | 1,891 |
| 3 | 2 | 3 | 2,548 |
| 4 | 4 | 6 | 2,793 |
| 5 | 7 | 11 | 3,437 |
| 6 | 12 | 17 | 2,927 |
| 7 | 18 | 26 | 3,306 |
| 8 | 27 | 41 | 3,378 |
| 9 | 42 | 70 | 3,262 |
| 10 | 71 | 711 | 3,330 |
| Avg(Basicness) Group | | | |
| 1 | 1.000 | 2.000 | 3,889 |
| 2 | 2.010 | 2.156 | 1,314 |
| 3 | 2.156 | 2.291 | 2,601 |
| 4 | 2.292 | 2.382 | 2,606 |
| 5 | 2.383 | 2.477 | 2,596 |
| 6 | 2.478 | 2.556 | 2,700 |
| 7 | 2.556 | 2.667 | 2,849 |
| 8 | 2.670 | 2.800 | 2,403 |
| 9 | 2.802 | 3.000 | 5,054 |
| Avg(Interdisciplinarity) Group | | | |
| 1 | 0.037 | 0.219 | 2,671 |
| 2 | 0.219 | 0.233 | 2,671 |
| 3 | 0.233 | 0.241 | 2,671 |
| 4 | 0.241 | 0.247 | 2,671 |
| 5 | 0.247 | 0.253 | 2,671 |
| 6 | 0.253 | 0.258 | 2,671 |
| 7 | 0.258 | 0.266 | 2,671 |
| 8 | 0.266 | 0.276 | 2,671 |
| 9 | 0.276 | 0.292 | 2,671 |
| 10 | 0.292 | 0.429 | 2,670 |
| Avg(Novelty) group | | | |
| 1 | 0.000 | 0.000 | 6,086 |
| 3 | 0.014 | 0.077 | 1,996 |
| 4 | 0.078 | 0.108 | 2,679 |
| 5 | 0.108 | 0.133 | 2,742 |
| 6 | 0.134 | 0.159 | 2,640 |
| 7 | 0.159 | 0.189 | 2,668 |
| 8 | 0.189 | 0.231 | 2,754 |
| 9 | 0.231 | 0.313 | 2,634 |
| 10 | 0.314 | 1.000 | 2,673 |
| Avg(Scientific citations) Group | | | |
| 1 | 0.000 | 16.846 | 2,688 |
| 2 | 16.857 | 31.036 | 2,687 |
| 3 | 31.053 | 47.407 | 2,687 |
| 4 | 47.417 | 65.000 | 2,691 |
| 5 | 65.015 | 83.738 | 2,683 |
| 6 | 83.750 | 104.630 | 2,688 |
| 7 | 104.638 | 131.128 | 2,687 |
| 8 | 131.128 | 166.667 | 2,688 |
| 9 | 166.672 | 243.296 | 2,686 |
| 10 | 243.400 | 5871.000 | 2,687 |

**Appendix II. Regression results based on front-page references.**

**Table A2. Front-page scientific references and patent citations**

|  | Patent citations | | | | | | |
|---|---|---|---|---|---|---|---|
|  | NB | | | | | | |
|  | (1) | (2) | (3) | (4) | (5) | (6) | (7) |
| I(sNPR) | 0.229*** | | | | | | |
|  | (0.035) | | | | | | |
| ln(sNPRs) |  | 0.210*** | 0.222*** | 0.213*** | 0.210*** | 0.203*** | 0.214*** |
|  |  | (0.010) | (0.010) | (0.010) | (0.010) | (0.011) | (0.011) |
| Avg(Basicness) |  |  | -0.007 |  |  |  | 0.029 |
|  |  |  | (0.031) |  |  |  | (0.030) |
| Avg(Interdisciplinarity) |  |  |  | 1.759*** |  |  | 2.038*** |
|  |  |  |  | (0.356) |  |  | (0.393) |
| Avg(Novelty) |  |  |  |  | 0.102 |  | -0.019 |
|  |  |  |  |  | (0.071) |  | (0.077) |
| ln(Avg(Scientific citations)+1) |  |  |  |  |  | 0.025+ | 0.051** |
|  |  |  |  |  |  | (0.014) | (0.015) |
| Issue year | Y | Y | Y | Y | Y | Y | Y |
| IPC class | Y | Y | Y | Y | Y | Y | Y |
| N | 33337 | 29110 | 27999 | 28982 | 29110 | 29110 | 27959 |
| BIC | 152144 | 132262 | 127270 | 131696 | 132269 | 132267 | 127043 |

This table repeats Table 2 (based on in-text references) but uses science measures based on front-page references instead. Robust standard errors in parentheses. *** p < .001, **p < .01, *p < .05, +p < .10.

**Table A3. Front-page scientific references and patent market value**

| | ln(Market value) | | | | | | |
|---|---|---|---|---|---|---|---|
| | OLS | | | | | | |
| | (1) | (2) | (3) | (4) | (5) | (6) | (7) |
| I(sNPR) | 0.581*** | | | | | | |
| | (0.083) | | | | | | |
| ln(sNPRs) | | 0.164*** | 0.149*** | 0.163*** | 0.164*** | 0.139*** | 0.128*** |
| | | (0.019) | (0.020) | (0.019) | (0.019) | (0.021) | (0.021) |
| Avg(Basicness) | | | -0.321*** | | | | -0.329*** |
| | | | (0.062) | | | | (0.062) |
| Avg(Interdisciplinarity) | | | | -1.175 | | | -0.508 |
| | | | | (0.713) | | | (0.788) |
| Avg(Novelty) | | | | | 0.008 | | -0.131 |
| | | | | | (0.147) | | (0.159) |
| ln(Avg(Scientific citations)+1) | | | | | | 0.078** | 0.067* |
| | | | | | | (0.027) | (0.029) |
| Issue year | Y | Y | Y | Y | Y | Y | Y |
| IPC class | Y | Y | Y | Y | Y | Y | Y |
| N | 7336 | 6435 | 6203 | 6408 | 6435 | 6435 | 6191 |
| R2 | 0.078 | 0.071 | 0.072 | 0.071 | 0.071 | 0.073 | 0.073 |
| BIC | 31045 | 26862 | 25729 | 26739 | 26853 | 26844 | 25699 |

This table repeats Table 3 (based on in-text references) but uses science measures based on front-page references instead. Robust standard errors in parentheses. *** $p < .001$, **$p < .01$, *$p < .05$, +$p < .10$.

Table A4. Front-page references and patent citations: A nonparametric approach

| | Patent citations | | | | |
|---|---|---|---|---|---|
| | NB | | | | |
| | (1) | (2) | (3) | (4) | (5) |
| Group | sNPRs | Basicness | Interdisciplinarity | Novelty | Scientific citations |
| 2 | -0.086+ | -0.062 | 0.250*** | / | -0.226*** |
| | (0.046) | (0.052) | (0.061) | | (0.055) |
| 3 | -0.078 | -0.068 | 0.179** | -0.225** | -0.159** |
| | (0.063) | (0.051) | (0.061) | (0.075) | (0.054) |
| 4 | -0.014 | -0.008 | 0.142* | -0.070 | -0.129* |
| | (0.049) | (0.051) | (0.059) | (0.050) | (0.055) |
| 5 | -0.048 | -0.046 | 0.166** | -0.114* | -0.105+ |
| | (0.049) | (0.052) | (0.061) | (0.051) | (0.055) |
| 6 | 0.188*** | 0.100* | 0.159** | -0.074 | -0.060 |
| | (0.051) | (0.051) | (0.058) | (0.049) | (0.068) |
| 7 | 0.314*** | 0.015 | 0.148* | -0.038 | -0.145** |
| | (0.057) | (0.050) | (0.068) | (0.048) | (0.055) |
| 8 | 0.336*** | -0.087 | 0.268*** | -0.004 | -0.049 |
| | (0.047) | (0.056) | (0.057) | (0.061) | (0.058) |
| 9 | 0.434*** | -0.034 | 0.234*** | -0.040 | -0.069 |
| | (0.046) | (0.051) | (0.059) | (0.047) | (0.059) |
| 10 | 0.818*** | / | 0.348*** | 0.010 | 0.035 |
| | (0.044) | | (0.061) | (0.046) | (0.060) |
| ln(sNPRs) | / | 0.221*** | 0.214*** | 0.229*** | 0.211*** |
| | | (0.012) | (0.010) | (0.013) | (0.011) |
| Issue year | Y | Y | Y | Y | Y |
| IPC class | Y | Y | Y | Y | Y |
| N | 33337 | 27999 | 28982 | 29110 | 29110 |
| BIC | 151346 | 127297 | 131734 | 132317 | 132290 |
| Chi2 | 735*** | 19* | 45*** | 14+ | 37*** |
| LR Chi2 | 200*** | 25*** | 34*** | 14+ | 48*** |

This table repeats Table 4 (based on in-text references) but uses science measures based on front-page references instead. LR Chi2 is 871*** if test the model Column 1 against the model in Table A2 Column 1. Robust standard errors in parentheses. *** $p < .001$, **$p < .01$, *$p < .05$, +$p < .10$.

Table A5. Front-page references and patent market value: A nonparametric approach

| | ln(Market value) OLS | | | | |
|---|---|---|---|---|---|
| | (1) | (2) | (3) | (4) | (5) |
| Group | sNPRs | Basicness | Interdisciplinarity | Novelty | Scientific citations |
| 2 | 0.158 | 0.128 | -0.096 | / | -0.312** |
| | (0.106) | (0.111) | (0.103) | | (0.120) |
| 3 | 0.453** | 0.003 | -0.011 | -0.011 | -0.083 |
| | (0.130) | (0.112) | (0.103) | (0.156) | (0.121) |
| 4 | 0.537*** | 0.065 | 0.174+ | -0.018 | -0.125 |
| | (0.105) | (0.111) | (0.104) | (0.096) | (0.121) |
| 5 | 0.535*** | -0.006 | -0.001 | -0.024 | 0.001 |
| | (0.109) | (0.115) | (0.108) | (0.105) | (0.118) |
| 6 | 0.562*** | -0.069 | -0.028 | 0.122 | -0.381** |
| | (0.110) | (0.115) | (0.105) | (0.101) | (0.121) |
| 7 | 0.721*** | -0.357** | 0.112 | 0.127 | -0.083 |
| | (0.105) | (0.116) | (0.106) | (0.098) | (0.119) |
| 8 | 0.790*** | -0.390** | -0.226* | 0.158 | 0.070 |
| | (0.103) | (0.125) | (0.109) | (0.098) | (0.118) |
| 9 | 0.763*** | -0.328** | -0.082 | 0.109 | -0.099 |
| | (0.104) | (0.105) | (0.110) | (0.099) | (0.116) |
| 10 | 0.880*** | / | -0.213+ | -0.044 | 0.177 |
| | (0.102) | | (0.118) | (0.099) | (0.118) |
| ln(sNPRs) | / | 0.150*** | 0.148*** | 0.147*** | 0.157*** |
| | | (0.022) | (0.020) | (0.026) | (0.021) |
| Issue year | Y | Y | Y | Y | Y |
| IPC class | Y | Y | Y | Y | Y |
| N | 7336 | 6203 | 6408 | 6435 | 6435 |
| R2 | 0.089 | 0.076 | 0.074 | 0.073 | 0.078 |
| BIC | 31017 | 25756 | 26770 | 26897 | 26870 |
| F | 15*** | 7*** | 3** | 1 | 5*** |
| LR Chi2 | 13 | 26*** | 21** | 9 | 36*** |

This table repeats Table 5 (based on in-text references) but uses science measures based on front-page references instead. LR Chi2 is 81*** if test the model Column 1 against the model in Table A3 Column 1. Robust standard errors in parentheses. *** $p < .001$, **$p < .01$, *$p < .05$, +$p < .10$.

**Appendix III. Robustness tests**

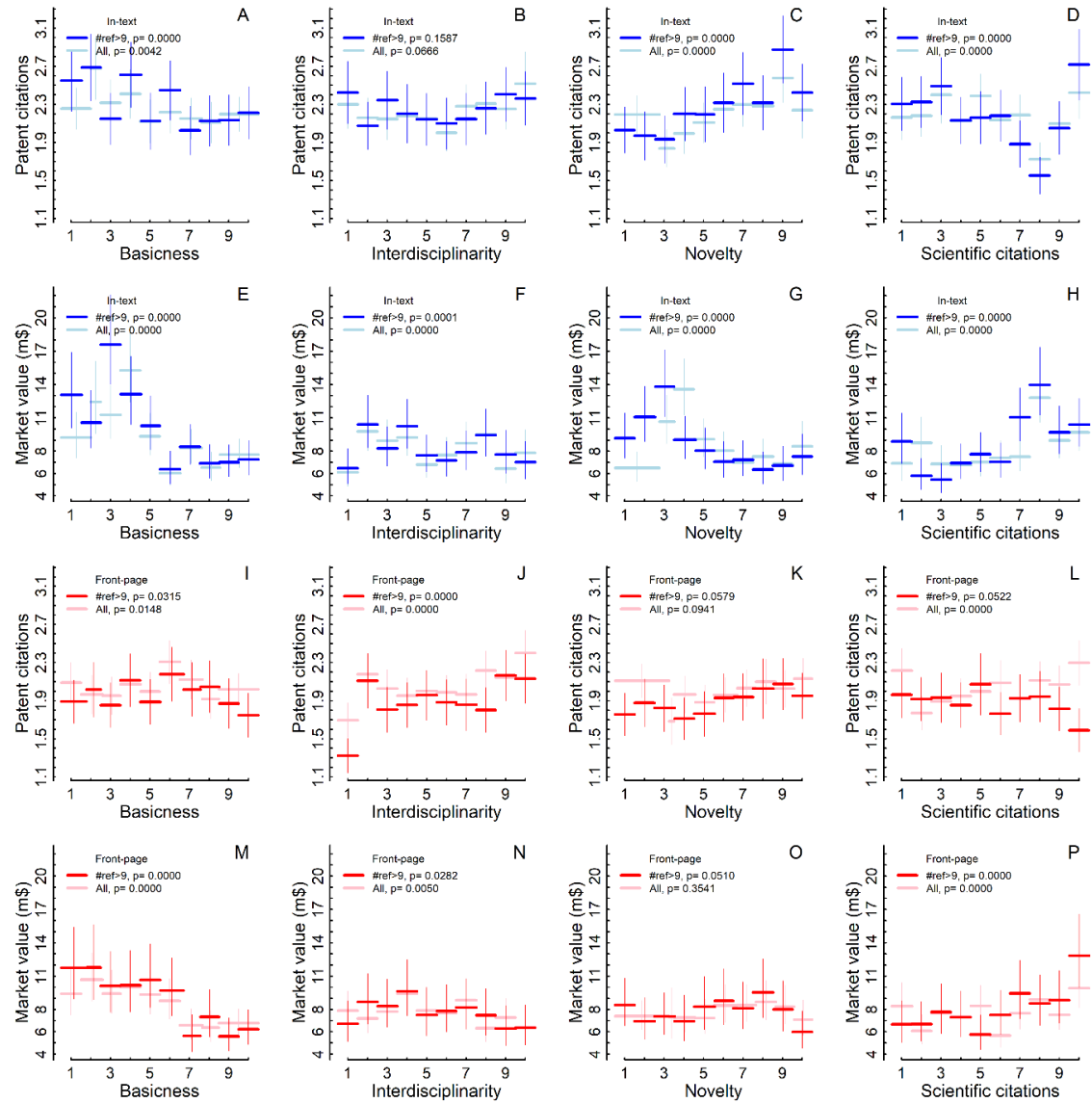

**Figure A1. Patents with at least 10 references.** Light blue lines are identical to the blue lines in Figure 3, and light red lines are identical to the red lines in Figure 4. We repeat the analyses using only patents with at least 10 scientific references. We categorize science measures within this subset of patents and then use them as independent variables. Results are plotted in blue and red lines.

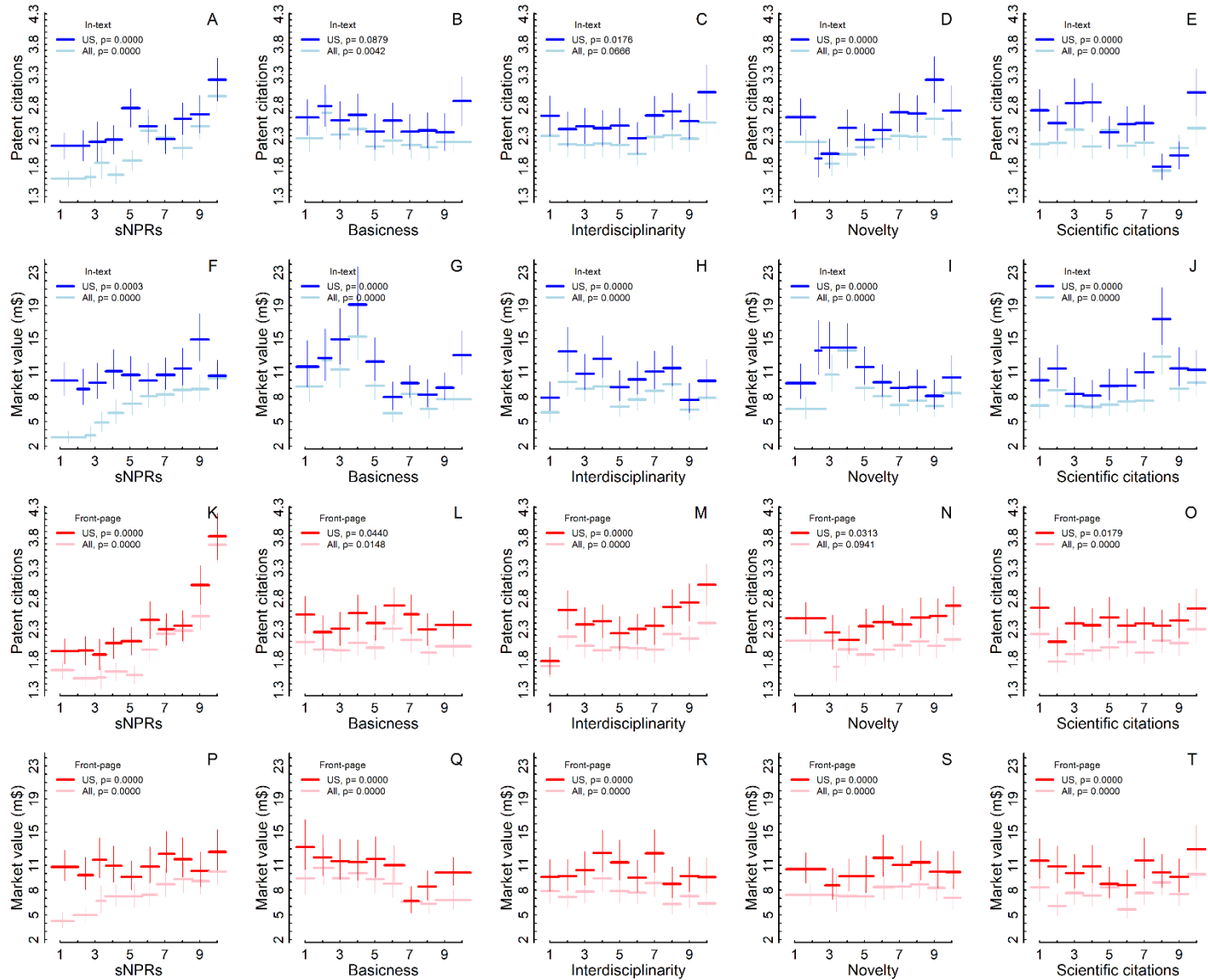

**Figure A2. Patents with US applicants and inventors.** Light blue lines are identical to the blue lines in Figure 3, and light red lines are identical to the red lines in Figure 4. We repeat the analyses using only patents with at least one applicant and at least one inventor located in the United States. We categorize science measures within this subset of patents and then use them as independent variables. Results are plotted in blue and red lines.